%% file: main.tex
\documentclass[conference]{IEEEtran}
\usepackage[utf8]{inputenc}
\usepackage[T1]{fontenc}
\usepackage[french,main=english]{babel}

\usepackage{cite}
\usepackage{amsmath,amssymb,amsfonts}
\usepackage{algorithmic}
\usepackage{textcomp}
\usepackage{xcolor}
\usepackage{verbatim}
\usepackage{eurosym}
\usepackage{amsmath}
\usepackage{nccmath}
\usepackage{mathrsfs}
\usepackage{mathtools, bm}
\usepackage{amssymb, bm}
\usepackage[ruled,vlined]{algorithm2e}
\usepackage{comment}

\usepackage{subcaption}

\usepackage{enumitem}
\usepackage{amsfonts}
\usepackage{relsize}
\usepackage{ragged2e}
\usepackage[free-standing-units=true]{siunitx}
\usepackage{url}

\usepackage{tikz}
\usetikzlibrary{positioning}
\usepackage{pgfplots}

\usepackage{graphicx} %%%%%%%%%%%%%%%%%%%%%%%%%%

\pgfplotsset{compat=newest}
\usetikzlibrary{shapes,arrows,fit,positioning,calc}
\usetikzlibrary{plotmarks}
\usetikzlibrary{decorations.pathreplacing}
\usetikzlibrary{patterns}
\usetikzlibrary{chains,arrows,shapes,spy}
\usetikzlibrary{automata}

\usepackage{textcomp}
\usetikzlibrary{shapes,arrows,backgrounds}
\usetikzlibrary{trees}

\usepackage{siunitx}

\usepackage{soul}
\usepackage{color}
\usepackage{ulem}

\def\BibTeX{{\rm B\kern-.05em{\sc i\kern-.025em b}\kern-.08em
    T\kern-.1667em\lower.7ex\hbox{E}\kern-.125emX}}

\begin{document}

\title{Physical Layer Authentication Using Information Reconciliation}

\author{\IEEEauthorblockN{Atsu Kokuvi Angélo Passah\IEEEauthorrefmark{1}\IEEEauthorrefmark{2},
Rodrigo C. de Lamare\IEEEauthorrefmark{1}\IEEEauthorrefmark{3}, and
Arsenia Chorti\IEEEauthorrefmark{2}\IEEEauthorrefmark{4}
}
\IEEEauthorblockA{\IEEEauthorrefmark{1}Department of Electrical Engineering (DEE), CETUC, PUC-Rio, Brazil}
\IEEEauthorblockA{\IEEEauthorrefmark{2}ETIS Laboratory UMR 8051, ENSEA, CY Cergy Paris University, CNRS, France \vspace{-0.2em}}
\IEEEauthorblockA{\IEEEauthorrefmark{3}School of Physics, Engineering and Technology, York University, United Kingdom}
\IEEEauthorblockA{\IEEEauthorrefmark{4}Barkhausen Institut gGmbH, Germany}
}

\maketitle

\begin{abstract}
User authentication in future wireless communication networks is expected to become more complicated due to their large scale and heterogeneity. Furthermore, the computational complexity of classical cryptographic approaches based on public key distribution can be a limiting factor for using in simple, low-end Internet of things (IoT) devices. This paper proposes physical layer authentication (PLA) expected to complement existing traditional approaches, e.g., in multi-factor authentication protocols. The precision and consistency of PLA is impacted because of random variations of wireless channel realizations between different time slots, which can impair authentication performance. In order to address this, a method based on error-correcting codes in the form of reconciliation is considered in this work. In particular, we adopt distributed source coding (Slepian-Wolf) reconciliation using polar codes to reconcile channel measurements spread in time. Hypothesis testing is then applied to the reconciled vectors to accept or reject the device as authenticated. Simulation results show that the proposed PLA using reconciliation outperforms prior schemes even in low signal-to-noise ratio scenarios.\\
\end{abstract}


% \begin{comment}
% \section*{Paper structure}
% \textcolor{blue}{
% I. Introduction: \\
% problem, motivation, prior work/literature, contributions \\
% \\, outline
% II. System model: \\
% system under consideration, notation and use of symbols, block diagram\\
% \\
% III. Proposed approach: \\
% main ideas, development of the technical solution, pseudo-code\\
% \\
% IV. Analysis:\\
% computational complexity, secrecy rate analysis etc\\
% \\
% V. Simulations/Numerical results: \\
% \\
% presentation of figures and discussion of results\\
% \\
% VI. Conclusions
% }
% \end{comment}

\input{Introduction}

\input{System_model}
\input{Proposed_approach}
\input{Analysis}

\input{Numerical_results}
\input{Conclusion}
\input{References}

\end{document}

%% file: Introduction.tex
\section{Introduction}
%\subsection{Motivations}
The emergence of new generation wireless systems such as large scale, heterogeneous, Internet of things(IoT) networks, brings about a lot of security issues. In this context, the computational complexity of classical cryptographic schemes that use public key encryption node authentication can be a limiting factor in terms of performance, inducing considerable delays\cite{ref1},\cite{ref2}. Furthermore, standard wireless communication systems do not employ security protocols at the physical layer. Physical layer security\cite{ref3} is currently being considered as the means to address such problems in next generation networks \cite{ref4} by incorporating them in new security protocols. Authentication is a key element in security protocols, in particular,  physical layer authentication (PLA) takes advantage of channel characteristics, similarly to hardware layer security, e.g., physical unclonable functions, that proposing authentication exploiting variations during hardware fabrication processes. %to protect the physical layer by preventing devices from impersonating attacks. % In fact, an impersonator can't access the channel features since they are decorrelated when the distance between the legitimate transmitter and the impersonator is above a half of wavelength. 

Recently, various channel-based PLA schemes have been studied. However, the wireless medium is subject to random variations over time. To account for this fact, the works in \cite{ref5}  and \cite{ref6} studied an authentication scheme based on the channel impulse response (CIR) while also integrating additional multipath delay characteristics of the wireless channel into the authentication framework. Additionally, the studies in \cite{ref5}  and \cite{ref6} used a two dimensional quantization method to mitigate random variations of amplitudes and delays. Moreover,  \cite{ref6} exploited the time-varying propagation delay to enhance authentication accuracy. 
In \cite{ref7}, the authors proposed two PLA schemes based on CIR without using a quantization algorithm in contrast with \cite{ref5} and \cite{ref6}. The idea was to avoid quantization errors introduced by the algorithm that could degrade the authentication performance. The first method used directly the estimated channel to make decisions about authentication, while the second approach exploited the channel correlation coefficient to enhance the performance.

To address the problem of channel state information (CSI) variations over time, we consider in this paper an approach based on error-correcting codes in the form of reconciliation. The proposed method employs a Slepian-Wolf coding scheme with polar codes that allows the reconciliation of discrepancies between different channel measurements in order to authenticate legitimate users. The authentication decision is therefore based on the comparison between reconciled vectors followed by hypothesis testing. In addition, we derive closed-form expressions of the probability distribution of the hypothesis test statistical variable and of the probability of false alarm and detection. 

The remainder of this paper is organized as follows. Section II presents the authentication system model and explains the authentication phases. In Section III, the proposed approach is described. 
%as well as a brief presentation of the prior approach in \hl{ref6}. 
We provide performance analyses in Section IV by deriving closed-form expressions of the probability of false alarm and detection. Then, simulation results are presented in Section V and the paper is concluded in Section VI.

% \hl{ref6}
% \hl{........}

% Channel based approaches\\
% Key based approaches\\
% milimetter wave MIMO\\
% RF fingerprinting\\
% ML based 
%\subsection*{Contributions}

%% file: System_model.tex
\section{System Model}

A standard wireless communication system of three nodes denoted by Alice, Bob and Eve is considered, where Alice and Bob are the legitimates nodes and Eve is an adversary. In this scheme, Bob wants to authenticate Alice while Eve is an active attacker that attempts to impersonate her. The objective is to design a scheme based on the CSI to distinguish Alice and Eve, both of which are equipped with a single antenna ($N_u = 1$, \text{where} $u\in \{a,e\}$. The indices $a$ and $e$ denote respectively Alice and Eve. Bob is equipped with $N_b$ antennas.

We assume that the communication takes place in a rich scattering environment so that channel characteristics of different users are spatially uncorrelated when the distance between them is greater than half a wavelength\cite{ref8}. Therefore, Alice's and Eve's estimated channels at Bob are uncorrelated. %\hl{Ref} 
The communication is divided into two steps: the training phase and the authentication phase, as shown in Fig. \ref{fig1}.

\subsubsection{Training phase}

This phase occurs offline. Bob estimates the Alice's CSI. More precisely, in time slot $t$, Bob estimates the CSI $\mathbf{h}_{a}(t) \in \mathbb{C}^{1 \times N_b}$ of Alice, where ${h_a}_{i} \sim \mathcal{C N}\left(0, \sigma_h^2\right)$, $i = 1, \ldots, N_b$, which will be used as a reference in the authentication phase to decide whether the user is Alice or not. %$\mathbf{h}_a(t) \in \mathbb{C}^{1 \times N_b}$, and %${h_a}_{i} \sim \mathcal{C N}\left(0, \sigma_h^2\right)$, $i = 1, \ldots, N_b$. 
%is a zero mean complex gaussian random variable
\subsubsection{Authentication phase}
During the second online phase, in a subsequent time slot $t+m$, Bob takes new CSI measurements of $\mathbf{h}_{u}(t+m) \in \mathbb{C}^{1 \times N_b}$, $u\in \{a,e\}$, that may either come from Alice or Eve and needs to make an authentication decision based on the CSI obtained in the training phase. Without loss of generality, we assume $m=1$. When the transmitted signal at $t+1$ comes from Eve, ${h_e}_{i}(t+1) \sim \mathcal{C N}\left(0, \sigma_h^2\right)$. We consider a scenario where the channel behaviour changes slowly over time. The channel between the same transmitter-receiver pair can then be well described by %When it comes from Alice, we assume that $\mathbf{h}_{a}(t+1)$follows%  
a first order Gauss-Markov process\cite{ref7}. Thus, the channel between Alice and Bob in the time slot $t+1$ is given by
\begin{equation}
\mathbf{h}_{a}(t+1) = \beta\mathbf{h}_{a}(t) + \sqrt{1-\beta^2}\mathbf{n}_a  %(t+1)   
\end{equation}
% \textcolor{red}{Angelo, can you check if there is a square in the second $\beta$ in (1)?}
where $\beta$ is the channel correlation coefficient and $\mathbf{n}_a$ is a measurement noise vector, ${n_a}_{i} \sim \mathcal{C N}\left(0, \sigma_h^2\right)$. The noise vector $\mathbf{n}_a$ is statistically independent of $\mathbf{h}_{a}$.

We are going to present in the following section, our proposed approach based on Slepian-Wolf decoding to reconcile discrepancies over the CSI measurements in time, depicted through the system model in Fig. \ref{fig1}.

%\hl{.......}

%% file: Proposed_approach.tex
\section{Proposed approach}
\begin{figure}[t]
     \centering
     \includegraphics[width=0.49\textwidth]{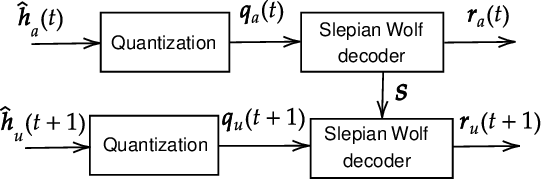}
      \caption{Proposed authentication scheme: the training phase at time $t$ and the authentication phase at time $t+1$, $u\in \{a,e\}$}
     \label{fig1}
\end{figure}

Our goal is to use reconciliation in order to reduce the impact of inconsistencies over the CSI observed in different time slots. Reconciliation is a standard technique used in physical unclonable function based authentication (typically referred to as fuzzy extractors) and secret key generation from channel measurements. It is proposed in this work, for the fist time, to be used in channel-based PLA. In each phase, the channel estimation is quantized and the output vectors at time $t$ and $t+1$ are treated as the codewords at the input of the reconciliation, as shown in Fig. \ref{fig1}. Note that during the offline first phase, not only the original CSI but also helper data (i.e., syndrome side information) need to be stored. Then, to make a decision during the online phase, a hypothesis test is performed by Bob in order to distinguish Alice from Eve. 
\subsubsection{Training phase} 
%\begin{itemize}
%\setlength{\itemsep}{6pt}
%\justifying
%    \item  Channel estimation\\
%\textcolor{red}{The notation for the channel estimates could be a bit different. For example, we could use $\hat{\mathbf{h}}_a(t)$ for the channel estimate.}
The measurement by Bob at time $t$ is given by
\begin{equation}
    \centering
\hat{\mathbf{h}}_{a}(t)=\mathbf{h}_{a}(t)+\mathbf{z}(t)
\label{equ2}
\end{equation}
where $\mathbf{z}(t) \in \mathbb{C}^{1 \times N_b}$ is a zero mean complex Gaussian noise so that $z_{i}(t) \sim \mathcal{C N}\left(0, \sigma_z^2\right)$, $i = 1, \ldots, N_b$. Bob collects $M$ samples of these measurements in a row vector $\left[ \hat{\mathbf{h}}_{a1}\|\hat{\mathbf{h}}_{a2}\| \ldots \|\hat{\mathbf{h}}_{aM} \right] \in \mathbb{C}^{1 \times MN_b}$. By extracting the real and imaginary parts of this row vector and by concatenating them, we get the vector $\mathbf{x}_a(t) \in \mathbb{R}^{1 \times N}$ where $N = 2MN_b$. 
% \begin{equation}  
%     \centering
% \mathbf{y}_{a}(t)=\mathbf{h}_{a}(t)+\mathbf{z}(t)
% \end{equation}
% where $\mathbf{z}(t) \in \mathbb{C}^{1 \times N_b}$ is a zero mean complex gaussian noise so that $z_{i}(t) \sim \mathcal{C N}\left(0, \sigma_z^2\right)$, $i = 1, \ldots, N_b$
%sss
%where $\mathbf{h}_a(t) \in \mathbb{C}^{1 \times N_aN_b}$, $h_{i} \sim \mathcal{C N}\left(0, \sigma_h^2\right)$, $i = 1, \ldots, N_aN_b$
%\item Quantization\\
%\textcolor{red}{Angelo, can you explain why we are using 1-bit quantization. In this context, why do not use extra bits for quantization? Please explain.}
For a purpose of simplicity, $\mathbf{x}_a(t)$ is then quantized using the 1-bit threshold quantizer (\ref{equ3}) and output $\mathbf{q}_a(t) \in \{0,1\}^{1 \times N}$. 
%The quanzation function is given by
\begin{equation}
    \centering
    \begin{aligned}
    Q(x_j)= \begin{cases}1, & \text { if } x_j \geq \gamma \\ 0, & \text { otherwise }\end{cases}
    \end{aligned}
    \label{equ3}
\end{equation}
where the threshold $\gamma$ is the mean of $\mathbf{x}$, i.e., $\gamma = {\rm mean}(\mathbf{x})$ 

\vspace{0.4cm}
\subsubsection{Authentication phase}
Similarly to the previous step, the channel measurements at Bob's side is given by
% \begin{itemize}
%     \item when it is Alice's channel
\begin{equation}
    \centering
\hat{\mathbf{h}}_{u}(t+1)=\mathbf{h}_{u}(t+1)+\mathbf{z}(t+1)
\label{equ4}
\end{equation}
where $u \in \{a,e\}$, $\mathbf{z}(t+1) \in \mathbb{C}^{1 \times N_b}$, $z_{i}(t+1) \sim \mathcal{C N}\left(0, \sigma_z^2\right)$, $i = 1, \ldots, N_b$, and the quantized vector is $\mathbf{q}_u(t+1) \in \{0,1\}^{1 \times N}$. Based on the principle of Slepian Wolf decoding with polar codes \cite{ref9}, $\mathbf{q}_a(t)$ and $\mathbf{q}_u(t+1)$ are decoded using the cyclic redundancy check ($\mathrm{CRC}$) successive cancellation list decoding. It enhances the performance of the polar code in finite blocklengths \cite{ref10},\cite{ref11}. $\mathrm{CRC}$ assists the decoder in choosing the right decoding path from a list of possibilities. The reconciliation outputs reconciled vectors $\mathbf{r}_a(t)$ and $\mathbf{r}_u(t+1)$ $\in  \left\{0,1\right\}^{1\times K}$. Note that the decoding of $\mathbf{q}_u(t+1)$ uses the syndrome $\mathbf{s}$ from the decoding of $\mathbf{q}_a(t)$. 
%and the first reconciled code $\mathbf{r}_a(t) \in  \left\{0,1\right\}^{1\times K}$.
%$\mathbf{s}$ and $\mathbf{r}_a(t) \in  \left\{0,1\right\}^{1\times K}$ are then stored in a database that represents the helper information Bob needs at the authentication phase to make decisions.
%\textcolor{red}{Angelo, the last sentence is not clear. Can you please check it and rewrite?}
%= \left[\mathbf{S}_1\|\mathbf{S}_2\right]  \in  \left\{0,1\right\}^{1\times (N-k)}$ and the first reconciled code $\mathbf{r}_1 \in  \left\{0,1\right\}^{1\times k}$ \\
%Bob then uses the syndrome $\mathbf{s}$ from the helper database to decode $\mathbf{q}_u(t+1)$ and gets $\mathbf{r}_u(t+1) \in \left\{0,1\right\}^{1\times K}$.
For a well-designed reconciliation scheme with the right choice of code rate, the reconciled vectors should be the same in the normal case and should be different in the spoofing case. Thus, in an ideal situation, Bob expects $\mathbf{r}_a(t) = \mathbf{r}_a(t+1)$ in the normal case and $\mathbf{r}_a(t) \neq \mathbf{r}_e(t+1)$ in the spoofing case. Bob uses a hypothesis test to distinguish the normal case and the spoofing case {(\ref{eq})}. $H_0$ denotes the normal case, i.e., the user is Alice at $t+1$, $H_1$ denotes the spoofing case, i.e., the user is not Alice, $\eta_{th}$ is the decision threshold and $\eta$ is the hypothesis test statistical variable.
\begin{equation}
    \centering
    \left\{ \begin{aligned}
%\begin{cases}
H_0: \hspace{0.2cm}&\eta=d\left(\mathbf{r}_a(t), \mathbf{r}_a(t+1)\right) \leq \eta_{t h} \\
%\vspace{0.2cm}
H_1: \hspace{0.2cm} &\eta=d\left(\mathbf{r}_a(t), \mathbf{r}_e(t+1)\right)>\eta_{t h}
    \end{aligned}
\right.
\label{eq}
\end{equation}
A suitable choice to distinguish Alice from Eve is to calculate the bit error, that is to determine the number of bit positions where $\mathbf{r}_a(t)$ and $\mathbf{r}_u(t+1)$ are different. That is well represented by the Hamming distance between $\mathbf{r}_a(t)$ and $\mathbf{r}_u(t+1)$. $\eta$ is therefore as follows,
\begin{equation}
\eta = d\left(\mathbf{r}_a(t), \mathbf{r}_u(t+1)\right)=\sum_{n=1}^K\left|r_{a,n}(t)-r_{u,n}(t+1)\right|
\label{eta}
\end{equation}
where $d(.,.)$ is the Hamming distance. 

%\subsection{Prior approaches}
%\subsection{Performance metrics}

To better position this work and highlight the merits of the proposed approach, let us present briefly other possible approaches in \cite{ref7} and \cite{ref12}. In \cite{ref7}, channel measurements $\hat{\mathbf{h}}_{a}(t)$ (\ref{equ2}) and $\hat{\mathbf{h}}_{u}(t+1)$ (\ref{equ4}) was considered for the hypothesis test as follows: 
\begin{equation}
    \centering
    \left\{ \begin{aligned}
%\begin{cases}
H_0: \hspace{0.2cm}&\eta_p \leq {\eta_p}_{t h} \\
%\vspace{0.2cm}
H_1: \hspace{0.2cm} &\eta_p >{\eta_p}_{t h}
%\end{cases}
    \end{aligned}
\right.
\end{equation}
where the statistical variable $\eta_p $ was given by the square of the $\ell_2$-norm between $\hat{\mathbf{h}}_{a}(t)$ and $\hat{\mathbf{h}}_{u}(t+1)$ as given by 
\begin{equation}
    \centering
\eta_p  =\sum_{i=1}^{N_{b}}\left(\hat{h}_{a,i}(t)-\hat{h}_{u,i}(t+1)\right)^2.
\end{equation}

A key-based PLA method was presented in \cite{ref12}. Note that key-based methods assumes that legitimate nodes Alice and Bob agreed on the same secret keys before the transmission. %and the communication scheme is different from the one in this paper. 
The steps are described as follows:
\begin{itemize}
    \item The user requests the authentication by sending data packets to Bob.
    \item Authentication inquiry:\\
    Bob generates then $\mathbf{s}_b=\left[s_{b, 1}, s_{b, 2}, \ldots, s_{b, N_b}\right]^{\mathrm{T}}$, $s_{b, i}=\exp \left(j \theta_{b, i}\right)$ and $\theta_{b,i}$ is uniformly distributed over $[0,2 \pi)$.
    The signal received by the user at time $t$ is $x_{u, i}(t)= h_{u, i}(t) s_{b, i}+z_{b, i}(t)$, $i = 1, \ldots, N_b$
    where $h_{u, i}(t) \sim \mathcal{C N}\left(0,\sigma_h^2\right)$ and $z_{b, i}(t) \sim \mathcal{C N}\left(0,\sigma_z^2\right)$.
    \item Authentication response:\\
    The user gets the phase $\theta_{u, i}$ of $x_{u, i}(t)$ and sends the response signal $\mathbf{s}_u=\left[s_{u, 1}, s_{u, 2}, \ldots, s_{u, N}\right]^{\mathrm{T}}$ to Bob, where $s_{u, i}=\exp \left\{j\left(\mathcal{M}\left(k_{u, i}\right)-\theta_{a, i}\right)\right\}$. $ \mathcal{M(\cdot)}$ is the secret key mapping function. It is given by: $\mathcal{M}\left(k_i\right)= 0$ if $k_i= \left[0 0\right]$, $\mathcal{M}\left(k_i\right)= \pi / 2$ if $k_i= \left[0 1\right]$, $\mathcal{M}\left(k_i\right)= \pi$ if $k_i= \left[1 1\right]$ and $\mathcal{M}\left(k_i\right)=   3 \pi / 2$ if $k_i= \left[1 0\right]$,
% \begin{equation}
% \centering
%     \mathcal{M}\left(k_i\right)= \begin{cases}0, & k_i=\left[\begin{array}{ll}
%     0 & 0
%     \end{array}\right] \\
%     \pi / 2, & k_i=\left[\begin{array}{ll}
%     0 & 1
%     \end{array}\right] \\
%     \pi, & k_i=\left[\begin{array}{ll}
%     1 & 1
%     \end{array}\right] \\
%     3 \pi / 2, & k_i=\left[\begin{array}{ll}
%     1 & 0
%     \end{array}\right],\end{cases}
% \end{equation}
where $k_i$ is the key bits. The shared key bits between the legitimate nodes are $k_{a,i} = k_{b,i}$.
\item Authentication completion:\\
Bob received a signal $x_{b, i}(t+1)=h_{u, i}(t+1) s_{u, i}+z_{b, i}(t+1)$ and gets his phase as previously. He then calculates 
$y_i = x_{b,i}(t+1)s_{b,i}$ to remove $\theta_{b,i}$. 
\end{itemize}
The hypothesis test is formulated as
\begin{equation}
\left\{\begin{array}{l}
H_0: \mathbf{K}_{t+1} = \mathbf{K}_B \\
H_1: \mathbf{K}_{t+1} \neq \mathbf{K}_B
\end{array}\right.
\end{equation}
The hypothesis test statistic variable ${\eta_p}\mathop {\stackrel {>} {\smash {\scriptstyle < }\vphantom {_{x}}}} \limits _{H_{0}}^{H_{1}}{\eta_p}_{t h}$ is given by
$
    \eta_p =\operatorname{Re}\left\{\sum\left(\exp ^{-j \mathcal{M}\left(\mathbf{K}_B\right)} \otimes \mathbf{y}\right)\right\}
$
where $\operatorname{Re}(\cdot)$ calculates the real part.

We focus in the next section on the analysis where we derive the closed-form expressions of the performance metrics.  %such as the probability of false alarm and the probability of detection.

%% file: Analysis.tex
\section{Performance Analysis}
Based on hypothesis testing, the performance metrics are the probability of false alarm (type I error) and the probability of detection. They are given respectively by $\mathrm{P_{FA}}=\operatorname{Pr}\left(\eta>\eta_{t h} \mid H_0\right)$ and $\mathrm{P_{D}}=\operatorname{Pr}\left(\eta > \eta_{t h} \mid H_1\right)$. Thus, we need to determine first the probability distribution of $\eta$ under $H_0$ and under $H_1$.

%\subsection{Probability distributions}
The behaviour of a Hamming distance (\ref{eta}) allows to determine the closed-form expressions of the probability distribution of $\eta$ under $H_0$ and $H_1$ in the following propositions.
\subsubsection*{Proposition 1}
Under $H_0$, $\eta$ follows a binomial distribution of parameters $K$ and $p_0$, i.e. $\eta \sim  \mathbb{B}(K,p_0)$.
\begin{equation}
    \centering
    P(\eta=n | H_0) = \binom{K}{n} p_0^n (1-p_0)^{K-n}
\end{equation}
where $p_0$ is the bit error probability during the decoding.
\subsubsection*{Proof} The Hamming distance $\eta$ can be modeled as the sum of Bernoulli random variables, where each bit has a certain probability of being in error. Let's consider $X_i$, a Bernoulli random variable representing the error in the $i^{th}$ bit position.
\begin{equation}
    \centering
    \left\{ \begin{aligned}
X_i = 1\hspace{0.2cm}&\text{,if error} \\
%\vspace{0.2cm}
X_i = 0 \hspace{0.2cm} &\text{,otherwise}
    \end{aligned}
\right.
\end{equation}
$\implies \eta = \sum_{i=1}^KX_i$. $X_i$, $i = 1, \ldots, K$ are independent and identically distributed $\implies$ $\eta$ follows a binomial distribution of parameters $K$ and $p_0$, $\eta \sim  \mathbb{B}(K,p_0)$, where $p_0$ is the bit error probability under the normal case $H_0$. The pdf of $\eta$ is then given by $P(\eta=n | H_0) = \binom{K}{n} p_0^n (1-p_0)^{K-n}$.

\subsubsection*{Proposition 2} Under $H_1$, $\eta$ follows a binomial distribution of parameters $K$ and $p_1$, i.e. $\eta \sim  \mathbb{B}(K,p_1)$. 
\begin{equation}
    \centering
P(\eta=n | H_1) = \binom{K}{n} p_1^n (1-p_1)^{K-n}
\end{equation}
where $p_1$ is the bit error probability during the decoding. 
\subsubsection*{Proof} It is very similar to that of Proposition 1 by replacing $p_0$ by $p_1$ $\implies P(\eta=n | H_1) = \binom{K}{n} p_1^n (1-p_1)^{K-n}$.

\begin{figure}[t]
     \centering
     \includegraphics[width=0.45\textwidth]{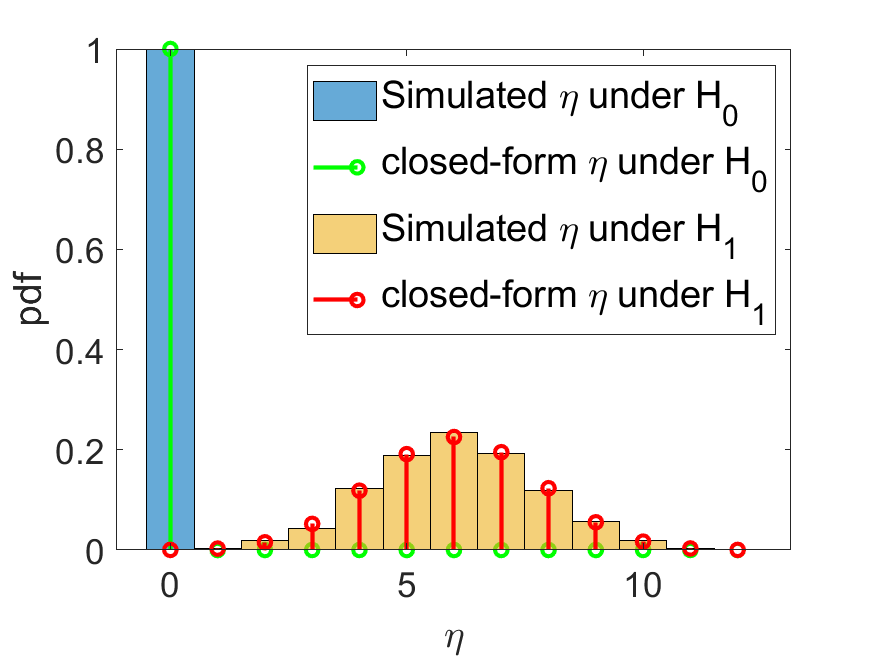}
     \caption{Simulated vs closed-from expression: code rate $ = 0.01$, $SNR = 10dB$, $p_0 \approx 0$ and $p_1 \approx 0.5025$}
     \label{fig2}
\end{figure}

In propositions 1 and 2, bit error probabilities $p_0$ and $p_1$ can be estimated with the bit error rate by simulation. Fig. \ref{fig2} shows the comparison between simulation results and closed-from expressions of the probability distribution of $\eta$ under $H_0$ and $H_1$. The closed-from expression perfectly matches the simulation result under both hypotheses for a $SNR = 10dB$.

Based on the proposition 1, the closed-form expression of the false alarm probability is given by
%and the detection probability are given as follows. 
%\subsubsection*{Proposition 3}
%The probability of false alarm is given by 
\begin{equation}
    \centering
 P_{FA} = \sum_{n=\eta_{th}+1}^{K} \binom{K}{n} p_0^n (1-p_0)^{K-n}. 
\end{equation}
%\subsubsection*{Proposition 4}
Based on the proposition 2, the closed-form expression of the  probability of detection is given by
%The probability of detection is given by
\begin{equation}
    \centering
 P_D = \sum_{n=\eta_{th}+1}^{K} \binom{K}{n} p_1^n (1-p_1)^{K-n}.
\end{equation}

%\subsubsection*{Proposition 3} The closed-form expression of the hypothesis test threshold $\eta$ given the false alarm probability $P_{FA} = \alpha$ is given by

% Therefore, the threshold $\eta_th$ can be determined given the probability of false alarm and then calculate the probability of detection in order to compare the simulated probabilities with the theoretical ones.

%\vspace{0.1cm}

%\subsection{Probability of false alarm and of detection}

%\subsection{Threshold}

%% file: Numerical_results.tex
\section{Numerical Results}
% \begin{figure}[h!]
%      \centering
%      \begin{subfigure}[b]{0.5\textwidth}
%          \centering
%          \includegraphics[width=\textwidth]{figures/probd005.png}
%          \caption{Code rate = 0.05}
%          %\label{fig:sw_cod}
%      \end{subfigure}
%      \begin{subfigure}[b]{0.5\textwidth}
%          \centering
%          \includegraphics[width=\textwidth]{figures/probd001.png}
%          \caption{Code rate = 0.01}
%          %\label{fig:sw_decod}
%      \end{subfigure}
%      \caption{Proposed authentication scheme}
% \end{figure}
% \begin{figure}[h!]
% \centering
% \input{figures/graphic1}
% \caption{Probability of detection vs Probability of false alarm}
% %\label{fig1}    
% \end{figure}
The simulation parameters are defined as follows. The number of antennas at the receiver, i.e., Bob, is $N_b = 32$, the channel correlation coefficient $\beta = 0.9$, the channel variance $\sigma_h^2 = 1$, $M = 16$, the codelength $N = 2MN_b = 1024$. The signal-to-noise ratio is defined as $SNR = \frac{\sigma_h^2}{\sigma_z^2}$. 

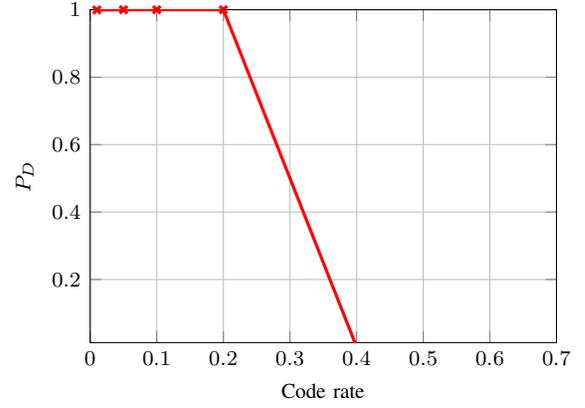
\begin{figure}[t]
\centering
\input{figures/PdvsRate}
\caption{$P_D$ vs code rate: $P_{FA} = 0.001$, $SNR = 15dB$}
\label{fig3}    
\end{figure}

First, we investigate the impact of the code rate on the reconciliation scheme. Fig. \ref{fig3} shows the detection probability as a function of the code rate for a $SNR = 15dB$ and a false alarm rate of $10^{-3}$. We observe that the detection probability is almost equal to $1$ for code rates less than $0.2$. For code rates greater than $0.2$, it becomes very low. In this case, as the code rate increases, it decreases towards $0$. The result can be improved in higher $SNR$ scenarios or by increasing the code length.  
% \begin{figure}[t]
% \centering
% \includegraphics[width=0.45\textwidth]{figures/pdvsrate.eps}
% \caption{$P_D$ vs code rate: $P_{FA} = 0.001$, $SNR = 15dB$}
% \label{fig3}    
% \end{figure}

\begin{figure}[t]
\centering
\input{figures/PdvsPfa}
\caption{ROC curve: code rate $= 0.01$, $SNR = 5dB$}
\label{fig4}    
\end{figure}
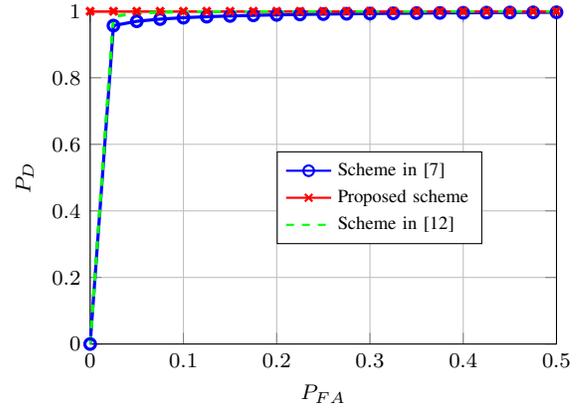

We compare our results with the prior results in \cite{ref7} and \cite{ref12}. In Fig. \ref{fig4}, we present the receiver operating characteristic (ROC) curve that represents the probability of detection as a function of the probability of false alarm for a $SNR = 5dB$ and a code rate of $0.01$. As the $P_{FA}$ increases, the $P_D$ increases for all the schemes. We observe that not only the proposed reconciliation method performs better than the prior ones but also $P_D$ is always very close to $1$ even for very low probabilities of detection. We actually have approximately $99.97\%$ increase in the detection probability for a $P_{FA} \approx 0\%$. That can be explained by the fact that using the error-correcting code allows to reconcile subsequent channel measurements. We also notice that the prior scheme in \cite{ref12} performs better than the one in \cite{ref7} with a small gap. % for lower $P_{FA}$. 
%\textcolor{red}{Angelo, can you please explain the existing schemes and also use the associated references? Can you also increase the fonts of the legends and the axes?}.
% \begin{figure}[t]
% \centering
% \includegraphics[width=0.45\textwidth]{figures/pdvspfa.eps}
% \caption{ROC curve: code rate $= 0.01$, $SNR = 5dB$}
% \label{fig4}    
% \end{figure}

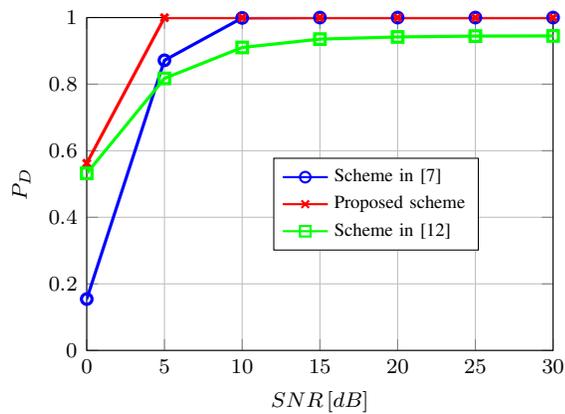
\begin{figure}[t]
\centering
\input{figures/PdvsSNR}
\caption{$P_D$ vs $SNR$: code rate $= 0.01$ with a $P_{FA} = 0.001$ }
\label{fig5}    
\end{figure}

Fig. \ref{fig5} presents the detection probability as a function of the SNR for a code rate of $0.01$ and a false alarm probability $P_{FA} = 10^{-3}$. First, as the SNR increases, the $P_D$ increases for all the schemes. This is due to the fact that, as the SNR increases, the estimation errors  decline. Second, our reconciliation scheme performs better than prior results even for low SNRs but the work in \cite{ref7}'s performance is very close for SNRs greater than $10dB$. We notice that the method in \cite{ref12}'s performance is better than the one in \cite{ref7} for SNRs less that $5dB$ but is worse than \cite{ref7} for SNRs greater than $5dB$. %Also, the gap with the prior scheme in \cite{ref7} is significant for SNRs under $5dB$. %For a $SNR \leq 5$, the performance of the proposed reconciliation scheme improves by increasing the codelength or the code rate as explained in Fig. \ref{fig3}. 
% \begin{figure}[t]
% \centering
% \includegraphics[width=0.45\textwidth]{figures/pdvssnr.eps}
% \caption{$P_D$ vs $SNR$: code rate $= 0.01$ with a $P_{FA} = 0.001$ }
% \label{fig5}    
% \end{figure}

%% file: figures/PdvsRate.tex
% This file was created by matlab2tikz v0.4.7 running on MATLAB 8.3.
% Copyright (c) 2008--2014, Nico Schlömer <nico.schloemer@gmail.com>
% All rights reserved.
% Minimal pgfplots version: 1.3
%
% The latest updates can be retrieved from
%   http://www.mathworks.com/matlabcentral/fileexchange/22022-matlab2tikz
% where you can also make suggestions and rate matlab2tikz.
%
%
% defining custom colors
\usetikzlibrary{positioning,calc}

\definecolor{mycolor1}{rgb}{0.00000,1.00000,1.00000}%
\definecolor{mycolor2}{rgb}{1.00000,0.00000,1.00000}%

\definecolor{mustard}{rgb}{0.92941,0.69020,0.12941}%

\definecolor{newpurple}{rgb}{0.5, 0 ,1}%

\definecolor{darkblue}{rgb}{0, 0.4470, 0.7410}

\pgfplotsset{every axis label/.append style={font=\footnotesize},
every tick label/.append style={font=\footnotesize},
every plot/.append style={ultra thick}
}

\begin{tikzpicture}[font=\footnotesize]

\begin{axis}[%
%tiny,
name=mse,
%ymode=log,
width  = 0.7\columnwidth,%5.63489583333333in, %0.55
%height = 0.3\columnwidth,%4.16838541666667in,
height = 0.5\columnwidth,%4.16838541666667in, %0.45
scale only axis,
xmin  = 0, %0.1,
xmax  = 0.7,  %1,
xlabel= {Code rate},
xmajorgrids,
ymin=0.013,  %0,  %0.05,
ymax=1,
ymode= linear, %normal = linear, %log,
ylabel={$P_{D}$},
ymajorgrids,
% legend entries={d-signature,
% $H_{v}$-signature,
%                % List-MMSE-Soft-IC,
% %MMSE based comparator network,
% },
% legend style={fill=white, fill opacity=0.6, draw opacity=1,
% text opacity =1,at={(0.4,0.3)}, anchor= south west,draw=black,fill=white,legend cell align=left,font=\scriptsize}
]

% \addlegendimage{smooth,color=black,solid, thick, mark=x,
% y filter/.code={\pgfmathparse{\pgfmathresult-0}\pgfmathresult}}
% \addlegendimage{smooth,color=red,solid, thick, mark=square,
% y filter/.code={\pgfmathparse{\pgfmathresult-0}\pgfmathresult}}
% \addlegendimage{smooth,color=green,solid, thick, mark=o,
% y filter/.code={\pgfmathparse{\pgfmathresult-0}\pgfmathresult}}

% Ellipsoids

% \draw (25,0.0025) ellipse (0.2cm and 0.4cm);
% \draw[dspconn]    (25,0.0043) -- (20,0.02) ;
% \draw (20,0.03) node [anchor=north west][inner sep=0.75pt]  [font=\footnotesize]  {\text{APs-Sel}};

% \draw (25,0.0003) ellipse (0.2cm and 0.5cm);
% \draw[dspconn]    (24,0.0002) -- (10,0.00025) ;
% \draw (4,0.0003) node [anchor=north west][inner sep=0.75pt]  [font=\footnotesize]  {\text{All-APs}};

         %smooth,
\addplot+[color=red,solid,thick, every mark/.append style={solid} ,mark=x,line width=1.2pt,
y filter/.code={\pgfmathparse{\pgfmathresult-0}\pgfmathresult}]
  table[row sep=crcr]{%
% 0.01  0.999352029718102\\
% 0.05  1\\
% 0.1   1\\
% 0.2   0.038588471977961\\
% 0.4   0.030664053820745\\
% 0.5   0.024647513701841\\
% 0.7   0.016819561354127\\
0.01  0.999210117338355\\
0.05  0.999999999999433\\
0.1   1\\
0.2   1\\
0.4   3.502799117823179e-05\\
0.5   2.290090238694407e-04\\
0.7   2.737436031962713e-04\\
};

\end{axis}
\end{tikzpicture}

%% file: figures/PdvsPfa.tex
% This file was created by matlab2tikz v0.4.7 running on MATLAB 8.3.
% Copyright (c) 2008--2014, Nico Schlömer <nico.schloemer@gmail.com>
% All rights reserved.
% Minimal pgfplots version: 1.3
%
% The latest updates can be retrieved from
%   http://www.mathworks.com/matlabcentral/fileexchange/22022-matlab2tikz
% where you can also make suggestions and rate matlab2tikz.
%
%
% defining custom colors
\usetikzlibrary{positioning,calc}

\definecolor{mycolor1}{rgb}{0.00000,1.00000,1.00000}%
\definecolor{mycolor2}{rgb}{1.00000,0.00000,1.00000}%

\definecolor{mustard}{rgb}{0.92941,0.69020,0.12941}%

\definecolor{newpurple}{rgb}{0.5, 0 ,1}%

\definecolor{darkblue}{rgb}{0, 0.4470, 0.7410}

\pgfplotsset{every axis label/.append style={font=\footnotesize},
every tick label/.append style={font=\footnotesize},
every plot/.append style={ultra thick}
}

\begin{tikzpicture}[font=\footnotesize]

\begin{axis}[%
%tiny,
name=mse,
%ymode=log,
width  = 0.7\columnwidth,%5.63489583333333in,
%height = 0.3\columnwidth,%4.16838541666667in,
height = 0.5\columnwidth,%4.16838541666667in,
scale only axis,
xmin  = 0, %0.1,
xmax  = 0.5, %1,
xlabel= {$P_{FA}$},
xmajorgrids,
ymin=0,  %0,  %0.05,
ymax=1,
ymode= linear, %normal = linear, %log,
ylabel={$P_{D}$},
ymajorgrids,
xtick={0, 0.1, 0.2, 0.3, 0.4, 0.5},
legend entries={Scheme in \cite{ref7},
Proposed scheme, Scheme in \cite{ref12}
               % List-MMSE-Soft-IC,
%MMSE based comparator network,
},
legend style={fill=white, fill opacity=0.6, draw opacity=1,
text opacity =1,at={(0.4,0.3)}, anchor= south west,draw=black,fill=white,legend cell align=left,font=\scriptsize}
]

\addlegendimage{smooth,color=blue,solid, thick, mark=o,
y filter/.code={\pgfmathparse{\pgfmathresult-0}\pgfmathresult}}
\addlegendimage{smooth,color=red,solid, thick, mark=x,
y filter/.code={\pgfmathparse{\pgfmathresult-0}\pgfmathresult}}
\addlegendimage{smooth,color=green,dashed, thick, mark=,
y filter/.code={\pgfmathparse{\pgfmathresult-0}\pgfmathresult}}

% Ellipsoids

% \draw (25,0.0025) ellipse (0.2cm and 0.4cm);
% \draw[dspconn]    (25,0.0043) -- (20,0.02) ;
% \draw (20,0.03) node [anchor=north west][inner sep=0.75pt]  [font=\footnotesize]  {\text{APs-Sel}};

% \draw (25,0.0003) ellipse (0.2cm and 0.5cm);
% \draw[dspconn]    (24,0.0002) -- (10,0.00025) ;
% \draw (4,0.0003) node [anchor=north west][inner sep=0.75pt]  [font=\footnotesize]  {\text{All-APs}};

%scheme in [7]
        %smooth,
\addplot+[color=blue,solid,thick, every mark/.append style={solid} ,mark=o,line width=1.2pt,
y filter/.code={\pgfmathparse{\pgfmathresult-0}\pgfmathresult}]
  table[row sep=crcr]{%
% 0    0\\          
% 0.0500  0.969612707753174\\
% 0.1000  0.980370638199819\\
% 0.1500  0.985635403605007\\
% 0.2000  0.988904984956173\\
% 0.2500  0.991173539438183\\
% 0.3000  0.992853173587351\\
% 0.3500  0.994151067992475\\
% 0.4000  0.995184692199065\\
% 0.4500  0.996026410636350\\
% 0.5000  0.996723601562157\\
% 0.5500  0.997308767776669\\
% 0.6000  0.997805047394408\\
% 0.6500  0.998229414601794\\
% 0.7000  0.998594637577612\\
% 0.7500  0.998910531451086\\
% 0.8000  0.999184799050895\\
% 0.8500  0.999423641998810\\
% 0.9000  0.999632322990951\\
% 0.9500  0.999816245840232\\
% 1.0000  1\\
  0   0\\
0.0250 0.956691163190838\\
0.0500 0.969612707753174\\
0.0750 0.976172884645796\\
0.1000 0.980370638199819\\
0.1250 0.983362799564267\\
0.1500 0.985635403605007\\
0.1750 0.987435607023703\\
0.2000 0.988904984956173\\
0.2250 0.990131577193882\\
0.2500 0.991173539438183\\
0.2750 0.992071110127261\\
0.3000 0.992853173587351\\
0.3250 0.993541093449987\\
0.3500 0.994151067992475\\
0.3750 0.994695638914494\\
0.4000 0.995184692199065\\
0.4250 0.995626142087458\\
0.4500 0.996026410636350\\
0.4750 0.996390771545565\\
0.5000 0.996723601562157\\
};

%\addlegendentry{MMSE APS}

%proposed scheme
        %smooth, 
\addplot+[color=red,solid, thick, every mark/.append style={solid} ,mark=x,line width=1.2pt,
y filter/.code={\pgfmathparse{\pgfmathresult-0}\pgfmathresult}]
  table[row sep=crcr]{%
%  0  0.999736502055099\\
% 0.0500  0.999736502055099\\
% 0.1000  0.999736502055099\\
% 0.1500  0.999736502055099\\
% 0.2000  0.999736502055099\\
% 0.2500  0.999736502055099\\
% 0.3000  0.999736502055099\\
% 0.3500  0.999736502055099\\
% 0.4000  0.999736502055099\\
% 0.4500  0.999736502055099\\
% 0.5000  0.999736502055099\\
% 0.5500  0.999736502055099\\
% 0.6000  0.999736502055099\\
% 0.6500  0.999736502055099\\
% 0.7000  0.999736502055099\\
% 0.7500  0.999736502055099\\
% 0.8000  0.999736502055099\\
% 0.8500  0.999736502055099\\
% 0.9000  0.999736502055099\\
% 0.9500  0.999736502055099\\
% 1.0000  0.999736502055099\\
 0 0.999736502055099\\
0.0250 0.999736502055099\\
0.0500 0.999736502055099\\
0.0750 0.999736502055099\\
0.1000 0.999736502055099\\
0.1250 0.999736502055099\\
0.1500 0.999736502055099\\
0.1750 0.999736502055099\\
0.2000 0.999736502055099\\
0.2250 0.999736502055099\\
0.2500 0.999736502055099\\
0.2750 0.999736502055099\\
0.3000 0.999736502055099\\
0.3250 0.999736502055099\\
0.3500 0.999736502055099\\
0.3750 0.999736502055099\\
0.4000 0.999736502055099\\
0.4250 0.999736502055099\\
0.4500 0.999736502055099\\
0.4750 0.999736502055099\\
0.5000 0.999736502055099\\
};
%\addlegendentry{SIC APS}

%scheme in [12]
        %smooth, 
\addplot+[color=green,dashed, thick, every mark/.append style={solid} ,mark=,line width=1.2pt,
y filter/.code={\pgfmathparse{\pgfmathresult-0}\pgfmathresult}]
  table[row sep=crcr]{%
% 0 0\\  
% 0.0500  0.993830490515590\\
% 0.1000  0.998156613418377\\
% 0.1500  0.999249970968975\\
% 0.2000  0.999650488791886\\
% 0.2500  0.999824659628203\\
% 0.3000  0.999908182918557\\
% 0.3500  0.999950743907394\\
% 0.4000  0.999973282749568\\
% 0.4500  0.999985500044546\\
% 0.5000  0.999992200348626\\
% 0.5500  0.999995881346016\\
% 0.6000  0.999997887839793\\
% 0.6500  0.999998961964928\\
% 0.7000  0.999999519777709\\
% 0.7500  0.999999796271396\\
% 0.8000  0.999999924008195\\
% 0.8500  0.999999976910705\\
% 0.9000  0.999999995162040\\
% 0.9500  0.999999999581652\\
% 1.0000  1\\
 0 0\\
0.0250 0.984382426524458\\
0.0500 0.993830490515590\\
0.0750 0.996825612372564\\
0.1000 0.998156613418377\\
0.1250 0.998851144012813\\
0.1500 0.999249970968975\\
0.1750 0.999494094488261\\
0.2000 0.999650488791886\\
0.2250 0.999754143013721\\
0.2500 0.999824659628203\\
0.2750 0.999873624891101\\
0.3000 0.999908182918557\\
0.3250 0.999932892076887\\
0.3500 0.999950743907394\\
0.3750 0.999963748331892\\
0.4000 0.999973282749568\\
0.4250 0.999980307096684\\
0.4500 0.999985500044546\\
0.4750 0.999989347345014\\
0.5000 0.999992200348626\\
};

\end{axis}
\end{tikzpicture}

%% file: figures/PdvsSNR.tex
% This file was created by matlab2tikz v0.4.7 running on MATLAB 8.3.
% Copyright (c) 2008--2014, Nico Schlömer <nico.schloemer@gmail.com>
% All rights reserved.
% Minimal pgfplots version: 1.3
%
% The latest updates can be retrieved from
%   http://www.mathworks.com/matlabcentral/fileexchange/22022-matlab2tikz
% where you can also make suggestions and rate matlab2tikz.
%
%
% defining custom colors
\usetikzlibrary{positioning,calc}

\definecolor{mycolor1}{rgb}{0.00000,1.00000,1.00000}%
\definecolor{mycolor2}{rgb}{1.00000,0.00000,1.00000}%

\definecolor{mustard}{rgb}{0.92941,0.69020,0.12941}%

\definecolor{newpurple}{rgb}{0.5, 0 ,1}%

\definecolor{darkblue}{rgb}{0, 0.4470, 0.7410}

\pgfplotsset{every axis label/.append style={font=\footnotesize},
every tick label/.append style={font=\footnotesize},
every plot/.append style={ultra thick}
}

\begin{tikzpicture}[font=\footnotesize]

\begin{axis}[%
%tiny,
name=mse,
%ymode=log,
width  = 0.7\columnwidth,%5.63489583333333in,
%height = 0.3\columnwidth,%4.16838541666667in,
height = 0.5\columnwidth,%4.16838541666667in,
scale only axis,
xmin  = 0, %0.1,
xmax  = 30,
xlabel= {$SNR \left[dB\right]$},
xmajorgrids,
ymin=0,  %0,  %0.05,
ymax=1,
ymode= linear, %normal = linear, %log,
ylabel={$P_{D}$},
ymajorgrids,
legend entries={Scheme in \cite{ref7},
Proposed scheme, Scheme in \cite{ref12}
               % List-MMSE-Soft-IC,
%MMSE based comparator network,
},
legend style={fill=white, fill opacity=0.6, draw opacity=1,
text opacity =1,at={(0.4,0.3)}, anchor= south west,draw=black,fill=white,legend cell align=left,font=\scriptsize}
]

\addlegendimage{smooth,color=blue,solid, thick, mark=o,
y filter/.code={\pgfmathparse{\pgfmathresult-0}\pgfmathresult}}
\addlegendimage{smooth,color=red,solid, thick, mark=x,
y filter/.code={\pgfmathparse{\pgfmathresult-0}\pgfmathresult}}
\addlegendimage{smooth,color=green,solid, thick, mark=square,
y filter/.code={\pgfmathparse{\pgfmathresult-0}\pgfmathresult}}

% Ellipsoids

% \draw (25,0.0025) ellipse (0.2cm and 0.4cm);
% \draw[dspconn]    (25,0.0043) -- (20,0.02) ;
% \draw (20,0.03) node [anchor=north west][inner sep=0.75pt]  [font=\footnotesize]  {\text{APs-Sel}};

% \draw (25,0.0003) ellipse (0.2cm and 0.5cm);
% \draw[dspconn]    (24,0.0002) -- (10,0.00025) ;
% \draw (4,0.0003) node [anchor=north west][inner sep=0.75pt]  [font=\footnotesize]  {\text{All-APs}};

%scheme in [7]
        %smooth,
\addplot+[color=blue,solid,thick, every mark/.append style={solid} ,mark=o,line width=1.2pt,
y filter/.code={\pgfmathparse{\pgfmathresult-0}\pgfmathresult}]
  table[row sep=crcr]{%
% 0  0.418099088278247\\
% 5  0.952084987837664\\
% 10  0.999637137409469\\
% 15  0.999992719077728\\
% 20  0.999998751745413\\
% 25  0.999999333920590\\
% 30  0.999999467273678\\
0  0.154347414166334\\
5  0.871564139318412\\
10  0.998999246315982\\
15  0.999984643477189\\
20  0.999997734698727\\
25  0.999998850540755\\
30  0.999999095270296\\
};

%\addlegendentry{MMSE APS}

%proposed scheme
        %smooth, 
\addplot+[color=red,solid, thick, every mark/.append style={solid} ,mark=x,line width=1.2pt,
y filter/.code={\pgfmathparse{\pgfmathresult-0}\pgfmathresult}]
  table[row sep=crcr]{%
% 0  0.878638890583343\\
% 5  0.999736502055099\\
% 10  0.999757524946881\\
% 15  0.999784732403993\\
% 20  0.999670523091755\\
% 25  0.999653752198308\\
% 30  0.999743939977045\\
0  0.562431502907286\\
5  0.999736502055099\\
10  0.999757524946881\\
15  0.999784732403993\\
20  0.999670523091755\\
25  0.999653752198308\\
30  0.999743939977045\\
};
%\addlegendentry{SIC APS}

%scheme in [12]
        %smooth, 
\addplot+[color=green,solid, thick, every mark/.append style={solid} ,mark=square,line width=1.2pt,
y filter/.code={\pgfmathparse{\pgfmathresult-0}\pgfmathresult}]
  table[row sep=crcr]{%
% 0  0.883192997440233\\
% 5  0.979838311348460\\
% 10  0.994051172593657\\
% 15  0.996505440935021\\
% 20  0.997064445974780\\
% 25  0.997272226394117\\
% 30  0.997319329260356\\
0  0.532160404366360\\
5  0.817263795666926\\
10  0.910215100659080\\
15  0.935301406361839\\
20  0.941803400961804\\
25  0.944362659262707\\
30  0.944933295732440\\
};

\end{axis}
\end{tikzpicture}

%% file: Conclusion.tex
\section{Conclusion}
This paper investigates the problem of PLA using error-correcting polar code to reconcile discrepancies between channel measurements and to distinguish between the normal case and the spoofing case. We provide closed-form expressions for the false alarm probability and the detection probability. In a scenario of $SNR = 15dB$ and a false alarm rate of $10^{-3}$, results show that that the probability of detection is close to one for code rates less than or equal to $0.2$ and very small for code rates greater that $0.2$. Simulation results confirm  also that our reconciliation-based method has better performance than prior schemes. The fact that we use a small value of $10^{-3}$ for the false alarm probability allows to confirm the performance of our work in practical systems that need very low false alarm probabilities.

\section*{Acknowledgment}
 Atsu Kokuvi Angélo Passah has been supported by the ELIOT ANR-18-CE40-0030 and FAPESP 2018/12579-7, FAPERJ project. Arsenia Chorti has been partially supported by the EC through the Horizon Europe/JU SNS project ROBUST-6G (Grant Agreement no. 101139068), the ANR-PEPR 5G Future Networks project, the ELIOT ANR-18-CE40-0030 and FAPESP 2018/12579-7, FAPERJ project and the CYU INEX-PHEBE projects. 

%% file: References.tex
%\section*{References}*
%\newpage
%\newpage
% 2)
%%%%%% L. Xiao, L. J. Greenstein, N. B. Mandayam and W. Trappe, "Channel-based spoofing detection in frequency-selective rayleigh channels," in IEEE Transactions on Wireless Communications, vol. 8, no. 12, pp. 5948-5956, December 2009, doi: 10.1109/TWC.2009.12.081544.
%%%%%%%A. Mukherjee, S. A. A. Fakoorian, J. Huang and A. L. Swindlehurst, "Principles of Physical Layer Security in Multiuser Wireless Networks: A Survey," in IEEE Communications Surveys and Tutorials, vol. 16, no. 3, pp. 1550-1573, Third Quarter 2014, doi: 10.1109/SURV.2014.012314.00178.
%%%%%%J. Liu and X. Wang, "Physical Layer Authentication Enhancement Using Two-Dimensional Channel Quantization," in IEEE Transactions on Wireless Communications, vol. 15, no. 6, pp. 4171-4182, June 2016, doi: 10.1109/TWC.2016.2535442.
%%%%%%%B. Ahuja, D. Mishra and R. Bose, "Fair Subcarrier Allocation for Securing OFDMA in IoT Against Full-Duplex Hybrid Attacker," in IEEE Transactions on Information Forensics and Security, vol. 16, pp. 2898-2911, 2021, doi: 10.1109/TIFS.2021.3067157.

% \vspace{12pt}
% \color{red}
% IEEE conference templates contain guidance text for composing and formatting conference papers. Please ensure that all template text is removed from your conference paper prior to submission to the conference. Failure to remove the template text from your paper may result in your paper not being published.

%% file: main.bbl
\begin{thebibliography}{00}
\bibitem{ref1}Shakiba-Herfeh, M., Chorti, A., Vincent Poor, H. (2021). \textit{Physical Layer Security: Authentication, Integrity, and Confidentiality.} In: Le, K.N. (eds) Physical Layer Security. Springer, Cham. %https://doi.org/10.1007/978-3-030-55366-1-6
\bibitem{ref2}M. Mitev, A. Chorti, H. V. Poor and G. P. Fettweis, "What Physical Layer Security Can Do for 6G Security," in \textit{IEEE Open Journal of Vehicular Technology,} vol. 4, pp. 375-388, 2023. %doi: 10.1109/OJVT.2023.3245071.
\bibitem{ref3}Bloch, M., and Barros, J. (2011). \textit{Physical-Layer Security: From Information Theory to Security Engineering.} Cambridge: Cambridge University Press. %doi:10.1017/CBO9780511977985
\bibitem{ref4}B. Aazhang et al., "Key drivers and research challenges for 6G ubiquitous wireless intelligence",(white paper), Sep. 2019
\bibitem{ref5}F. J. Liu, X. Wang and S. L. Primak, "A two dimensional quantization algorithm for CIR-based physical layer authentication," 2013 \textit{IEEE International Conference on Communications (ICC)}, Budapest, Hungary, 2013, pp. 4724-4728. % doi: 10.1109/ICC.2013.6655319.
\bibitem{ref6}J. Liu and X. Wang, "Physical Layer Authentication Enhancement Using Two-Dimensional Channel Quantization," in \textit{IEEE Transactions on Wireless Communications,} vol. 15, no. 6, pp. 4171-4182, June 2016. % doi: 10.1109/TWC.2016.2535442.
\bibitem{ref7}N. Xie, J. Chen and L. Huang, "Physical-Layer Authentication Using Multiple Channel-Based Features," in \textit{IEEE Transactions on Information Forensics and Security,} vol. 16, pp. 2356-2366, 2021. %, doi: 10.1109/TIFS.2021.3054534.
%%%%P. Zhang, Y. Shen, X. Jiang and B. Wu, "Physical Layer Authentication Jointly Utilizing Channel and Phase Noise in MIMO Systems," in IEEE Transactions on Communications, vol. 68, no. 4, pp. 2446-2458, April 2020, doi: 10.1109/TCOMM.2020.2967393.
\bibitem{ref8}Jakes William C. \textit{Microwave Mobile Communications.} IEEE, 1994. %DOI.org (Crossref), https://doi.org/10.1109/9780470545287.
\bibitem{ref9}E. Arikan, "Source polarization," \textit{2010 IEEE International Symposium on Information Theory,} Austin, TX, USA, 2010, pp. 899-903. %doi: 10.1109/ISIT.2010.5513567.
\bibitem{ref10}I. Tal and A. Vardy, "List decoding of polar codes," \textit{2011 IEEE International Symposium on Information Theory Proceedings,} St. Petersburg, Russia, 2011, pp. 1-5. % doi: 10.1109/ISIT.2011.6033904.
\bibitem{ref11}M. Shakiba-Herfeh and A. Chorti, "Comparison of Short Blocklength Slepian-Wolf Coding for Key Reconciliation," \textit{2021 IEEE Statistical Signal Processing Workshop (SSP),} Rio de Janeiro, Brazil, 2021, pp. 111-115. %doi: 10.1109/SSP49050.2021.9513785.
\bibitem{ref12}X. Lu, J. Lei, Y. Shi and W. Li, "Physical-Layer Authentication Based on Channel Phase Responses for Multi-Carriers Transmission," in \textit{IEEE Transactions on Information Forensics and Security,} vol. 18, pp. 1734-1748, 2023. %doi: 10.1109/TIFS.2023.3251093.
% \bibitem{ref13} X. Wu and Z. Yang, "Physical-Layer Authentication for Multi-Carrier Transmission," in IEEE Communications Letters, vol. 19, no. 1, pp. 74-77, Jan. 2015, doi: 10.1109/LCOMM.2014.2375191.
% \bibitem{ref14}J. Choi, "A Coding Approach With Key-Channel Randomization for Physical-Layer Authentication," in IEEE Transactions on Information Forensics and Security, vol. 14, no. 1, pp. 175-185, Jan. 2019, doi: 10.1109/TIFS.2018.2847659.
\end{thebibliography}
